\documentclass[Physsubmission, Phys]{SciPost}
\hypersetup{
    colorlinks,
    linkcolor={red!50!black},
    citecolor={blue!50!black},
    urlcolor={blue!80!black}
}

\usepackage[bitstream-charter]{mathdesign}
\usepackage{subcaption}
\usepackage{physics}

\makeatletter
\newsavebox{\@brx}
\newcommand{\llangle}[1][]{\savebox{\@brx}{\(\m@th{#1\langle}\)}%
  \mathopen{\copy\@brx\kern-0.5\wd\@brx\usebox{\@brx}}}
\newcommand{\rrangle}[1][]{\savebox{\@brx}{\(\m@th{#1\rangle}\)}%
  \mathclose{\copy\@brx\kern-0.5\wd\@brx\usebox{\@brx}}}
\makeatother

\DeclareSymbolFont{usualmathcal}{OMS}{cmsy}{m}{n}
\DeclareSymbolFontAlphabet{\mathcal}{usualmathcal}

\newcommand*\pFqskip{8mu}
\catcode`,\active
\newcommand*\pFq{\begingroup
        \catcode`\,\active
        \def ,{\mskip\pFqskip\relax}%
        \dopFq
}
\catcode`\,12
\def\dopFq#1#2#3#4#5{%
        {}_{#1}\phi_{#2}\biggl(\genfrac..{0pt}{}{#3}{#4}\Big|#5\biggr)%
        \endgroup
}

\begin{document}

% TODO: write your article's title here.
% The article title is centered, Large boldface, and should fit in two lines
\begin{center}{\Large \textbf{
Computation of entanglement entropy in inhomogeneous free fermions chains by algebraic Bethe ansatz\\
}}\end{center}

% TODO: write the author list here. Use initials + surname format.
% Separate subsequent authors by a comma, omit comma at the end of the list.
% Mark the corresponding author with a superscript *.
\begin{center}
Pierre-Antoine Bernard\textsuperscript{1$\star$}, Gauvain Carcone\textsuperscript{1}, Nicolas Crampé\textsuperscript{2} and
Luc vinet\textsuperscript{1,3}
\end{center}

% TODO: write all affiliations here.
% Format: institute, city, country
\begin{center}
{\bf 1} Centre de recherches mathématiques, Université de Montréal, P.O. Box 6128, Centre-ville
Station, Montréal (Québec), H3C 3J7, Canada
\\
{\bf 2} Institut Denis-Poisson CNRS/UMR 7013 - Université de Tours - Université d’Orléans, Parc de
Grandmont, 37200 Tours, France
\\
{\bf 3} IVADO, 6666 Rue Saint-Urbain, Montréal (Québec), H2S 3H1, Canada
\\
% TODO: provide email address of corresponding author
* bernardpierreantoine@outlook.com
\end{center}

\begin{center}
\today
\end{center}

% For convenience during refereeing (optional),
% you can turn on line numbers by uncommenting the next line:
%\linenumbers
% You should run LaTeX twice in order for the line numbers to appear.

\definecolor{palegray}{gray}{0.95}
\begin{center}
\colorbox{palegray}{
  \begin{tabular}{rr}
  \begin{minipage}{0.1\textwidth}
    \includegraphics[width=20mm]{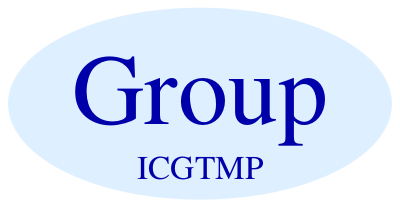}
  \end{minipage}
  &
  \begin{minipage}{0.85\textwidth}
    \begin{center}
    {\it 34th International Colloquium on Group Theoretical Methods in Physics}\\
    {\it Strasbourg, 18-22 July 2022} \\
    \doi{10.21468/SciPostPhysProc.?}\\
    \end{center}
  \end{minipage}
\end{tabular}
}
\end{center}

\section*{Abstract}
{\bf
The computation of the entanglement entropy for inhomogeneous free fermions chains based on $q$-Racah polynomials is considered. The eigenvalues of the truncated correlation matrix are obtained from the diagonalization of the associated Heun operator via the algebraic Bethe ansatz. In the special case of chains based on dual $q$-Hahn polynomials, the eigenvectors and eigenvalues are expressed in terms of symmetric polynomials evaluated on the Bethe roots.  
}

% TODO: include a table of contents (optional)
% Guideline: if your paper is longer that 6 pages, include a TOC
% To remove the TOC, simply cut the following block
\vspace{10pt}
\noindent\rule{\textwidth}{1pt}
\tableofcontents\thispagestyle{fancy}
\noindent\rule{\textwidth}{1pt}
\vspace{10pt}

\section{Introduction}
\label{sec:intro}

The characterization of entanglement in many-body systems is motivated by its numerous applications in quantum information \cite{amico2008entanglement, nielsen2002quantum} and its role in describing quantum critical points \cite{vidal2003entanglement}. This endeavour is usually carried out in bipartite situations, where the amount of entanglement between a region and its complement is determined. While many techniques have been developed to perform this task \cite{eisert2010colloquium}, analytical results for entanglement entropy in large systems remain rare. 
 
For spin chains and free fermions models, this problem reduces to diagonalizing a matrix referred to as the truncated correlation matrix \cite{Peschel_2003}. In cases where couplings are homogeneous, for example the XX spin chain, this matrix is Toeplitz or Toeplitz+Hankel and one can use the Fisher--Hartwig conjecture to compute the bipartite entanglement in the thermodynamic limit \cite{jin2004quantum, fagotti2011universal}. For more general couplings and truncated correlation matrices, applying these techniques is not possible and different approaches are required. 

Inhomogeneous fermionic chains associated to hypergeometric orthogonal polynomials of the Askey-Wilson scheme \cite{koekoek2010hypergeometric} are solvable and describe a wide variety of models. It was observed that their truncated correlation matrix admits a commuting tridiagonal matrix, identified as a Heun-Askey-Wilson operator \cite{gioev2006entanglement, eisler2013free, crampe2019free, FerBis, bernard2022entanglement}. This suggests an interesting connection with the theory of integrable systems.
Indeed, these operators arise in the transfer matrices associated to solutions of the reflection equations \cite{baseilhac2019diagonalization}. They correspond to Hamiltonians of XXZ spin chains with specific boundary fields and have been shown to be dagonalizable via the algebraic Bethe anstaz \cite{cao2003exact} (other methods have been developed in \cite{nepomechie2003bethe, crampe2010eigenvectors}). They are also examples of the so-called homogeneous case in the context of the modified algebraic Bethe ansatz, which has been designed
to deal with generic Heun operators \cite{ bernard2022bethe, bernard2021heun, baseilhac2019diagonalization} and diagonalize integrable models with arbitrary boundary conditions (see e.g. \cite{avan2015modified, belliard2013heisenberg, belliard2018modified, bernard2022bethe, crampe2017algebraic, crampe2015algebraic}). 

This paper applies the algebraic Bethe ansatz framework to investigate the spectrum of truncated correlation matrices of models associated to polynomials of the Askey-Wilson scheme. In particular, the eigenvalues of the truncated correlation matrix of free fermionic chains associated to dual q-Hahn polynomials are provided in terms of solutions of a set of Bethe equations. In section \ref{s2}, we recall the definition of free fermions chains associated to $q$-Racah polynomials and diagonalize their Hamiltonians. In section \ref{s3}, we discuss the problem of computing the entanglement entropy and introduce the truncated correlation matrix. In section \ref{s4}, we exhibit a commuting tridiagonal matrix referred to as the algebraic Heun operator and diagonalize it via the algebraic Bethe ansatz. This yields a set of relations known as Bethe equations. The eigenvalues of the truncated correlation matrix are then given in terms of roots of these equations. The associated $TQ$-relation and the thermodynamic limit are briefly discussed in section \ref{s5}.

\section{The model}
\label{s2}

\begin{figure}[h]
\captionsetup[subfigure]{justification=centering}
\begin{subfigure}{.5\textwidth}
  \centering
  \includegraphics[scale =0.5]{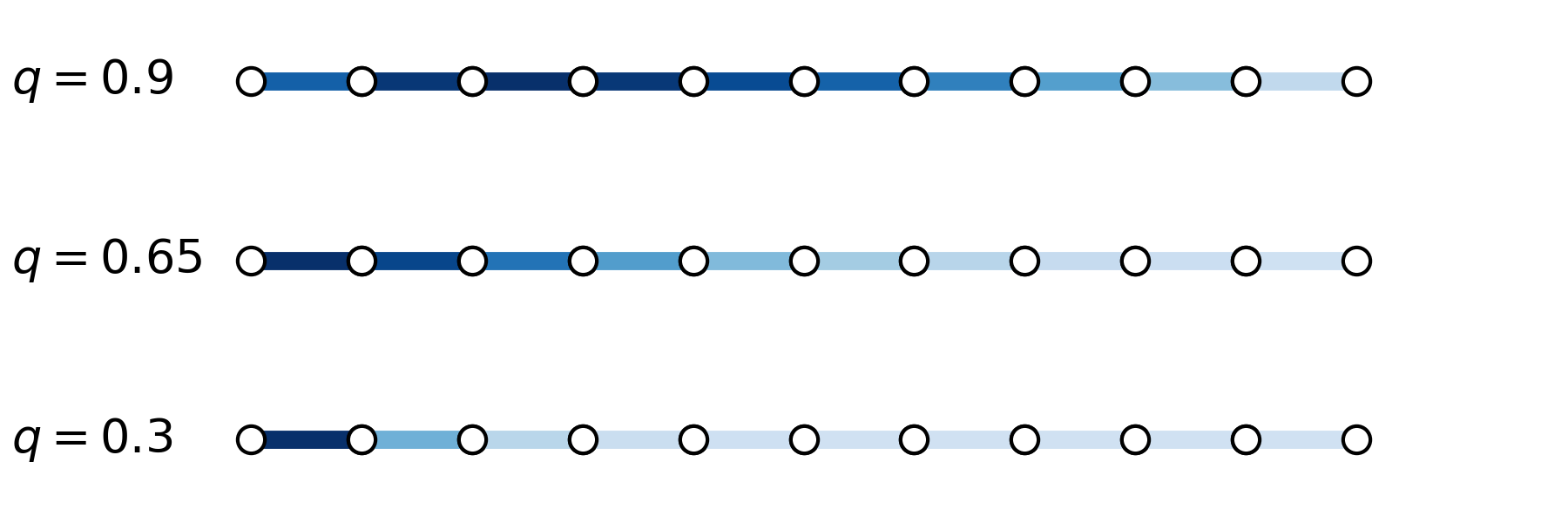}
  \caption{$\alpha = q^{-N-1},\ \beta = q^{2N}$,\\ $\gamma = q^{-2N},\ \delta = {(q^{-2N}  + q^{-N} )}/2 $}
  \label{fig:sfig1}
\end{subfigure}%
\begin{subfigure}{.5\textwidth}
  \centering
  \includegraphics[scale =0.5]{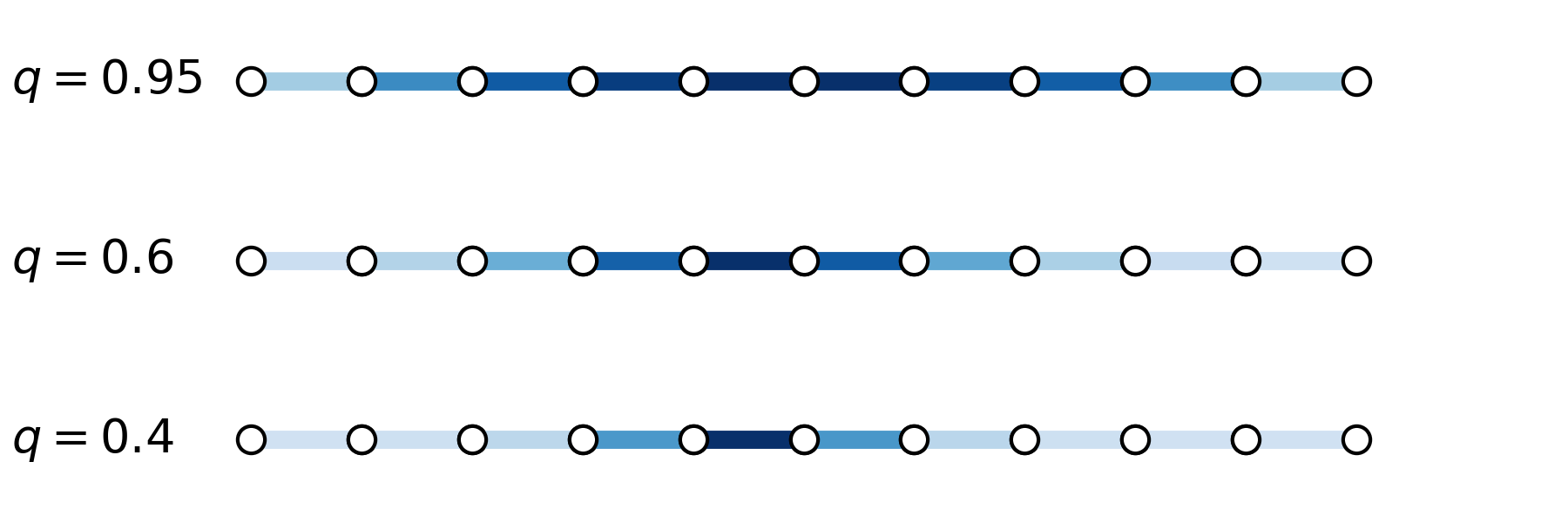}
  \caption{$\alpha = q^{-N-1},\ \beta = -q$, \\ $ \gamma = q^2/2,\  \delta = q^2/2 $}
  \label{fig:sfig2}
\end{subfigure}
\begin{subfigure}{.5\textwidth}
  \centering
  \includegraphics[scale =0.5]{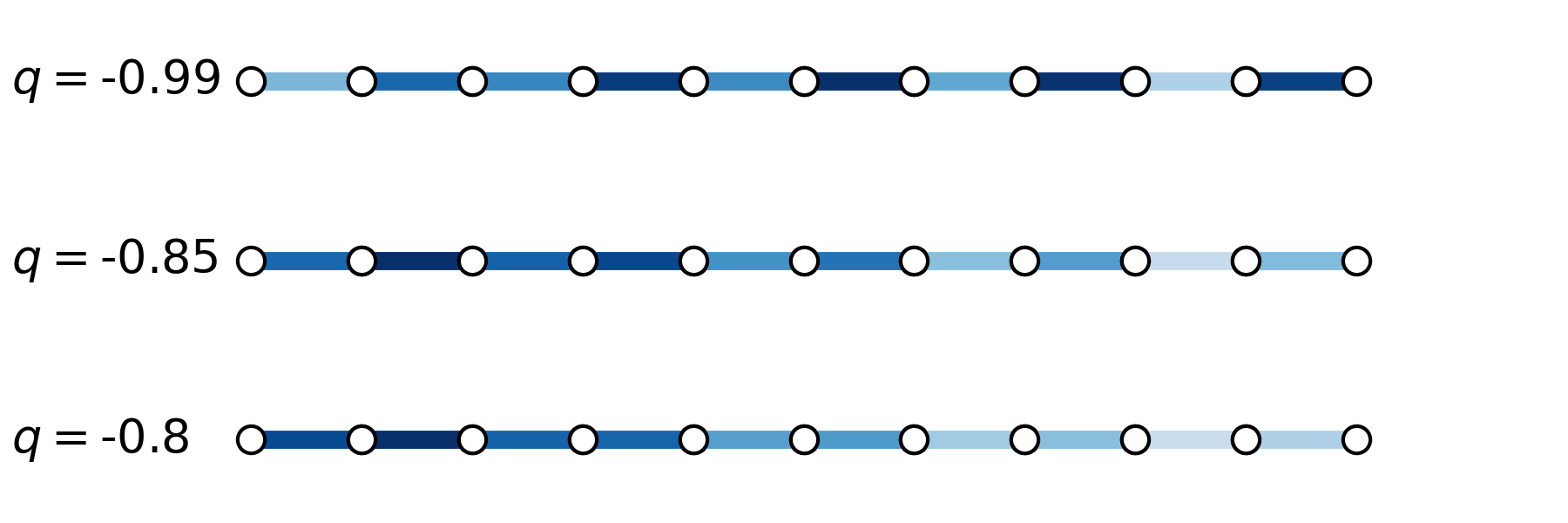}
  \caption{$\alpha = q^{-N-1}, \ \beta = q^{2N}$, \\ $\gamma = q^{-2N}, \ \delta = q^{-N-1} $ }
  \label{fig:sfig2}
\end{subfigure}
\begin{subfigure}{.5\textwidth}
  \centering
  \includegraphics[scale =0.5]{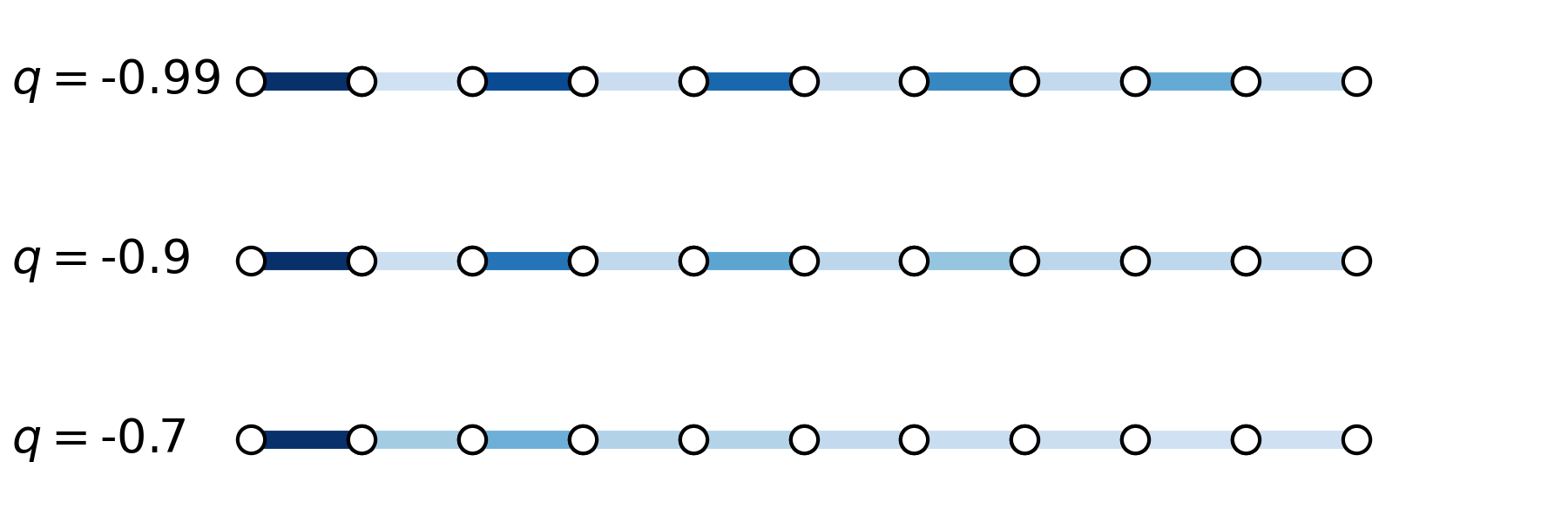}
  \caption{$\alpha = q^{-N-1}, \ \beta = q^{8N}$,\\ $ \gamma = q^{-2N}, \ \delta = q^{-8N} $}
  \label{fig:sfig2}
\end{subfigure}
\caption{Inhomogeneous free fermions chains of length $N= 10$, based on $q$-Racah polynomials, for different parameters $(q, \alpha, \beta, \gamma, \delta)$. The vertices and edges represent respectively the sites and the couplings. The color of the edges indicates the magnitude of $J_n$, \textit{i.e.} the strength of these couplings. Darker color is associated to stronger couplings. }
\label{fig1}
\end{figure}

Let us consider the following free fermions inhomogeneous Hamiltonian with nearest-neighbour interaction $J_n$ and magnetic field $\mu_n$,
\begin{equation}\label{eq:Hff}
\widehat{\mathcal{H}}=\sum_{n=0}^{N-1}(J_{n}  c_n^\dagger c_{n+1} + J_n c_{n+1}^{\dagger} c_{n})
- \sum_{n=0}^{N}\mu_n c_{n}^{\dagger} c_{n},
\end{equation}
where $c_n$ and $c^\dagger_n$ are fermionic annihilation and creation operators satisfying
\begin{equation}
     \{ c_{m}^{\dagger} 
\,, c^\dagger_{n} \} =  \{ c_{m} 
\,, c_{n} \} = 0, \quad  \{ c_{m}^{\dagger} 
\,, c_{n} \} = \delta_{m,n}.
\label{cr1}
\end{equation}
For convenience, we enumerate the sites of the lattice from $0$ to $N$. This model is equivalent to an inhomogeneous XX spin chain. Indeed, the Jordan-Wigner transformation
\begin{align}
    c_n^\dagger = \sigma^z_0 \sigma^z_1 \dots  \sigma^z_{n-1} \sigma^+_n, \quad \quad  c_n = \sigma^z_0 \sigma^z_1 \dots  \sigma^z_{n-1} \sigma^-_n,
\end{align}
allows to rewrite the canonical relations of the creation and annihilation operators \eqref{cr1} and the Hamiltonian \eqref{eq:Hff} in terms of spin-$1/2$ operators,
\begin{equation}
\widehat{\mathcal{H}}= -\frac{1}{2}\sum_{n=0}^{N-1}J_{n} ( \sigma^x_n \sigma^x_{n+1} + \sigma^y_n \sigma^y_{n+1} )
- \frac{1}{2}\sum_{n=0}^{N}\mu_n(1 + \sigma^z_n).
\end{equation}
We are interested in the case where the coupling parameters $J_n$ and the local magnetic field $\mu_n$ are constructed from the recurrence coefficients of the $q$-Racah polynomials \cite{koekoek2010hypergeometric}:
\begin{eqnarray}
    J_n&=& \epsilon  \sqrt{A_{n}C_{n+1} }\,,\\
    \mu_n&=& A_n+C_n -1 - \gamma\delta q  \,,
\end{eqnarray}
where $\epsilon=\pm1$ and $A_n,\ C_n$ are defined by
\begin{eqnarray}\label{eq:A}
    A_n&=& \frac{\left(\alpha  q^{n+1}-1\right)\left(\gamma  q^{n+1}-1\right) \left(\alpha  \beta  q^{n+1}-1\right) \left(\beta  \delta q^{n+1}-1\right)}
   {(1-\alpha  \beta  q^{2 n+1})(1-\alpha  \beta  q^{2 n+2})}\\
   \label{eq:C}
    C_n&=& \frac{ \left(\beta 
   q^n-1\right) \left(\alpha  q^n-\delta \right)
   \left(\alpha  \beta  q^n-\gamma \right)(q^{
   n+1}-q)}
   {(1-\alpha  \beta  q^{2 n})(1-\alpha 
   \beta  q^{2 n+1})}.
\end{eqnarray}
The choice of such inhomogeneous interactions and magnetic fields yields analytical results for the spectrum, as shown below. It also describes a large class of models thanks to the presence of various parameters. Indeed, the constants $A_n$, $C_n$ and $\mu_n$ depend on the five real parameters $q$, $\alpha$, $\beta$, $\gamma$ and $\delta$, restricted only by the requirement that
\begin{equation}
    A_n C_{n+1} > 0, \quad \quad J_{N} = 0.
    \label{condi1}
\end{equation}
For instance, as shown in figure \ref{fig1}, we can get couplings $J_n$ which are monotone in $n$ or peaking at a certain value. Taking $q<0$ also gives models with oscillating couplings, reminiscent of alternating spin chains \cite{totsuka1997magnetization}. 

\subsection{Diagonalization of the Hamiltonian}
In order to diagonalize $\widehat{\mathcal{H}}$, 
it is convenient to rewrite this operator in matrix form as
\begin{equation}\label{eq:Hff2}
\widehat{\mathcal{H}}=(c^\dagger_0,\dots,c^\dagger_N)\, \boldsymbol{A} 
\begin{pmatrix}
c_0 \\
\vdots \\
c_N
\end{pmatrix}
,
\end{equation}
where $\boldsymbol{A}$ is the hermitian $(N+1)\times (N+1)$ tridiagonal matrix given by
\begin{equation}\label{eq:Hh}
 \boldsymbol{A} =\sum_{n=0}^N\Big(J_{n} |n\rangle \langle n+1|  - \mu_n |n\rangle\langle n|  + J_{n}
 |n+1\rangle\langle n|\Big)\,, 
\end{equation}
with the convention $J_{N}=J_{-1}=0$. The vectors $\{|0\rangle,|1\rangle,\dots ,|N\rangle \}$ are naturally associated to sites in the chain and
give the canonical orthonormal basis of $\mathbb{C}^{N+1}$. They will be referred to as elements of the position basis. The spectral problem for $\boldsymbol{A}$ reads
\begin{equation}
\boldsymbol{A}  |\omega_k\rangle = \omega_k |\omega_k\rangle\ ,
\end{equation}
where
\begin{equation}
 |\omega_k\rangle =\sum_{n=0}^N \phi_n(\omega_k) |n\rangle\ . \label{eq:omm}
\end{equation}
Knowing that the entries of $\boldsymbol{A}$ are the recurrence coefficients of the $q$-Racah polynomials, one deduces (see eq. \eqref{recurel}) that
its eigenvalues $\omega_k$ are
\begin{equation}
    \omega_k=q^{-k}+\gamma\delta q^{k+1}.
\end{equation}
The wavefunctions $\phi_n(\omega_k) = \bra{\omega_k}\ket{n}$ are given in terms of $q$-Racah polynomials $R_n(\omega_k)$ \cite{koekoek2010hypergeometric}:
\begin{equation}
    \phi_n(\omega_k) = \epsilon^n\sqrt{W_{k}}\prod_{j=1}^n \sqrt{\frac{A_{j-1}}{C_{j}}}  R_n(\omega_k).
    \label{ovlap}
\end{equation}
The definition of $R_n(\omega_k)$ and the normalisation factors $W_{k}$ are given in appendix \ref{AppA}. The latter are chosen such that the wavefunctions $\phi_n(\omega_k)$ are orthonormal \textit{i.e.}
\begin{equation}
    \sum_{k=0}^N \phi_n(\omega_k)\phi_m(\omega_k)=\delta_{n,m} \quad \text{and} \quad  \sum_{n=0}^N \phi_n(\omega_k)\phi_n(\omega_{k'})=\delta_{k,k'}.
\end{equation}
From these wavefunctions, we can define new pairs of fermionic creation and annihilation operators in terms of which the Hamiltonian is diagonal:
\begin{equation}
    \widehat{\mathcal{H}} = \sum_{k = 0}^N \omega_k \Tilde{c}^\dagger_k \Tilde{c}_k, \quad \  \  k \in \{0, 1, \dots, N\}
\end{equation}
where
\begin{equation}
    \Tilde{c}_k = \sum_{n = 0}^N\phi_n(\omega_k) c_n,  \quad   \Tilde{c}^\dagger_k = \sum_{n = 0}^N\phi_n(\omega_k) c_n^\dagger.
\end{equation}
Note that the operators $\Tilde{c}_k^\dagger$ and $c_k$ are associated to the single particle excitations of the system, with energies given by the spectrum of the matrix $\boldsymbol{A}$. One may further observe that these energies are invariant under arbitrary transformations of $\alpha$, $\beta$ and under
\begin{equation}
    \delta \rightarrow \delta \kappa, \quad \gamma \rightarrow \gamma \kappa^{-1}, \quad \kappa \in \mathbb{R}.
\end{equation}
This is not true of the coupling parameters $J_n$ and local magnetic field $\mu_n$ which depend non-trivially on $(q,\alpha,\beta, \gamma, \delta)$. Important properties characterizing these systems, like the entanglement entropy in the ground state, should thus depend on these parameters. 

\section{Entanglement entropy}
\label{s3}
Entanglement in a multipartite system $A \cup B$ is measured by the entanglement entropy $S_A$, defined as
\begin{equation}
    S_A = - \text{tr}_A( \rho_A \ln{\rho_A}),
\end{equation}
where $A$ is a subsystem of $A \cup B$ with a reduced density matrix $\rho_A$ given by the trace over the degrees of freedom in $B$,
\begin{equation}
    \rho_A = \text{tr}_B |\Omega\rrangle \llangle \Omega|.
\end{equation}
In the following, we take $A$ to be the first $L+1$ sites of the inhomogeneous free fermionic chain introduced in the previous section. The states considered are obtained by filling up the first $K+1$ single particle states, taken as the Fermi sea,
\begin{equation}
    |\Omega\rrangle = \prod_{ k = 0}^K \Tilde{c}^\dagger_k |0\rrangle,
\end{equation}
where $|0\rrangle$ is the vacuum state annihilated by all operators $\Tilde{c}_k$. For $\omega_k$ monotone in $k$, $|\Omega\rrangle$ describes the ground state of Hamiltonians obtained as affine transformations of \eqref{eq:Hff2}. In other words, it gives the state for which the single particle excitations with negative energy are filled. \\

 As observed in \cite{Peschel_2003}, computing the entanglement entropy $S_A$ of free fermions can be done by diagonalizing the truncated correlation matrix. Indeed, it is known that \cite{carrasco2017duality}
\begin{equation}
    S_A = - \sum_{\ell} c_\ell \ln{c_\ell} + (1 -c_\ell) \ln{(1 - c_\ell)},
\end{equation}
where the coefficients $c_ \ell$ are the eigenvalues of the $(L+1) \times (L+1)$ matrix $C$ with entries $C_{nm}$ given by the $2$-point correlation functions,
\begin{equation}
 C_{nm} = \llangle \Omega | c_n^{\dagger} c_m |\Omega \rrangle =  \sum_{k=0}^K \phi_n(\omega_k) \phi_m(\omega_k), \quad n, m \in \{0, 1, \dots L+1\}.
 \label{Cposb}
\end{equation}
This is a submatrix of the complete correlation matrix of the ground state $\widehat{C}$, i.e.
\begin{equation}
    C = \pi_A \widehat{C} \pi_A, \quad \widehat{C} = \sum_{k = 0}^K\ket{\omega_k}\bra{\omega_k},
    \label{relationC}
\end{equation}
where $\pi_A$ is the projector onto the vector space associated to sites of subsystem $A$,
\begin{equation}
    \pi_A = \sum_{n = 0}^{L} \ket{n}\bra{n}.
\end{equation}
The computation of the entanglement entropy is thus reduced to determining the eigenvalues $c_\ell$. With the help of the algebraic Heun operators, we shall see that the spectral problem for the truncated correlation matrix can be treated in the algebraic Bethe ansatz framework.

\section{Algebraic Heun operator}
\label{s4}
In this section, we introduce a tridiagonal matrix that commutes with the truncated correlation matrix. To do so, we define an operator $\boldsymbol{A}^*$ which is diagonal in the position basis
\begin{equation}
    \boldsymbol{A}^* \ket{n} = \lambda_n \ket{n}, \quad \lambda_n = q^{-n} + \alpha \beta q^{n+1}.
    \label{ac1}
\end{equation}
Using the difference relation of the $q$-Racah polynomials and the expression \eqref{ovlap}, one finds the tridiagonal action of $\boldsymbol{A}^*$ on the eigenbasis of $\boldsymbol{A}$,
\begin{equation}
    \boldsymbol{A}^* \ket{\omega_k} = \Bar{J}_{k}  \ket{\omega_{k+1}}  - \Bar{\mu}_k \ket{\omega_k} +  \Bar{J}_{k-1}  \ket{\omega_{k-1}}.
    \label{tri2}
\end{equation}
The coefficients $\Bar{J}_{k}$ and $\Bar{\mu}_k$ are given in appendix \ref{AppA}. The Heun operator $T$ is then defined as
\begin{equation}
    T = \{\boldsymbol{A},\boldsymbol{A}^*\} - (\lambda_L + \lambda_{L+1}) \boldsymbol{A} - (\omega_K + \omega_{K+1}) \boldsymbol{A}^*,
\label{Hd1}
\end{equation}
and has the property of commuting with both the projector $\pi_A$ and the complete correlation matrix $\widehat{C}$,
\begin{equation}
    [T, \pi_A] = [T, \widehat{C}] = 0.
\end{equation}
This is shown easily by considering the commutators $[T, \pi_A]$ in the position basis and  $[T, \widehat{C}]$ in the energy basis. Given relation \eqref{relationC}, $T$ also commutes with the truncated correlation $C$ and thus share with it a common set of eigenvectors. This is a crucial observation, in particular because the Heun operator $T$ can be identified in the transfer matrix of integrable models and can hence be diagonalized via the algebraic Bethe ansatz \cite{baseilhac2019diagonalization,bergeron2020heun}.

\subsection{Algebraic Bethe ansatz}

The matrices $\boldsymbol{A}$ and $\boldsymbol{A}^*$ give a representation of the Askey-Wilson algebra \cite{crampe2021askey, zhedanov1991hidden}:

\begin{equation}
    \boldsymbol{A} \boldsymbol{A} \boldsymbol{A}^* - \left(q + \frac{1}{q}\right) \boldsymbol{A} \boldsymbol{A}^* \boldsymbol{A} +  \boldsymbol{A}^* \boldsymbol{A} \boldsymbol{A} = \xi \boldsymbol{A}  + \chi \boldsymbol{A}^* + \eta \mathcal{I},
    \label{aw1}
\end{equation}
\begin{equation}
    \boldsymbol{A}^* \boldsymbol{A}^* \boldsymbol{A} - \left(q + \frac{1}{q}\right) \boldsymbol{A}^* \boldsymbol{A} \boldsymbol{A}^* +  \boldsymbol{A} \boldsymbol{A}^* \boldsymbol{A}^* = \chi^* \boldsymbol{A} + \xi \boldsymbol{A}^* + \eta^*\mathcal{I} ,
    \label{aw2}
\end{equation}
where $\mathcal{I}$ is the $N+1\times N+1$ identity matrix and the constants $\xi$, $\chi$, $\chi^*$, $\eta$ and $\eta^*$ can be expressed in terms of the parameters in the Hamiltonian:
\begin{equation}
    \chi = -\frac{\gamma  \delta  \left(q^2-1\right)^2}{q}, \quad   \chi^* = - \frac{\alpha  \beta  \left(q^2-1\right)^2}{q},
\end{equation}
\begin{equation}
    \xi = -(q-1)^2 (\alpha  (\beta  \delta +\beta +\gamma +1)+\gamma  (\beta  \delta +\delta +1)+\beta  \delta ),
\end{equation}
\begin{equation}
    \eta =  (q-1)^2 (q+1) (\alpha  \gamma  (\beta  \delta +\delta +1)+\alpha  \beta  \delta +\gamma  \delta  (\beta  \delta +\beta +\gamma +1)),
\end{equation}
\begin{equation}
    \eta^* = (q-1)^2 (q+1) \left(\alpha ^2 \beta +\alpha  \left(\beta ^2 \delta +\beta  (\gamma +1) (\delta +1)+\gamma \right)+\beta  \gamma  \delta \right).
\end{equation}
The so-called dynamical operators can be defined in terms of the generators of this algebra:
\begin{equation}
\begin{split}
    \mathcal{A}(u,m) &= \frac{q^{-2L}}{\left(\alpha  \beta  q^{2 m+1}-1\right)} \left( \frac{q^{m+1}\{\boldsymbol{A},\boldsymbol{A}^*\}}{(q+1)}  - { \left(\alpha  \beta  q^{2m+2}+1\right)}\boldsymbol{A}  - \frac{ \left(q^{2m+2}+\gamma  \delta  u^4 \right)}{u^2 }  \boldsymbol{A}^* \right) + f_1(u,m) \mathcal{I},
\end{split}
\end{equation}
and
\begin{equation}
\begin{split}
       \mathcal{B}(u,m) &= \frac{\alpha  \beta   q^{m+2}+ q^{-m-1}}{2(q+1)} \{\boldsymbol{A},\boldsymbol{A}^*\}  -\frac{ q^{-m-1}-\alpha  \beta   q^{m+2}}{2(1-q)} [\boldsymbol{A},\boldsymbol{A}^*] \\
       &  -\alpha  \beta  (q+1)  \boldsymbol{A}  -\frac{ \left(q+\alpha  \beta  \gamma  \delta  u^4\right)}{u^2} \boldsymbol{A}^* + f_2(u,m) \mathcal{I}.
       \label{dyop1}
\end{split}
\end{equation}
The functions $f_1(u,m)$ and $f_2(u,m)$ are given in the appendix. These operators verify
\begin{equation}
     \mathcal{B}(u,m+1) \mathcal{B}(v,m) = \mathcal{B}(v,m+1) \mathcal{B}(u,m),
     \label{dor1}
\end{equation}
\begin{equation}
\begin{split}
     \mathcal{A}(u,m+1) \mathcal{B}(v,m) &=  f(u,v) \mathcal{B}(v,m) \mathcal{A}(u,m) + g(u,v,m) \mathcal{B}(u,m)\mathcal{A}(v,m) \\ & + w(u,v,m) \mathcal{B}(u,m) \mathcal{A}(\tau v^{-1},m),
     \end{split}
     \label{dor2}
\end{equation}
where $\tau = \sqrt{\frac{q}{\alpha \beta \gamma \delta}}$. The functions $f(u,v)$, $g(u,v,m)$ and $w(u,v,m)$ are given in appendix \ref{AppA}. Relations \eqref{dor1}-\eqref{dor2} were verified using directly the Askey-Wilson relations \eqref{aw1}-\eqref{aw2}. This is similar to the method used in \cite{bernard2021heun,bernard2022bethe} and distinct from the approach based on $R$ and $K$ matrices \cite{cao2003exact, baseilhac2019diagonalization}. The Heun operator \eqref{Hd1} can be expressed in terms of $\mathcal{A}(u,L)$ and $\mathcal{A}( \tau u^{-1},L)$ as
\begin{equation}
    T = r(u) \mathcal{A}(u,L) + r(\tau u^{-1}) \mathcal{A}(\tau u^{-1},L) - \left(r(u) f_1(u,L) + r\left(\tau u^{-1}\right) f_1\left(\tau u^{-1},L\right)\right)\mathcal{I},
    \label{be4}
\end{equation}
where
\begin{equation}
    r(u) = \frac{q^{L}(q+1)}{\alpha  \beta  \gamma  \delta  u^4-q}\left( \alpha ^2 \beta ^2 \gamma  \delta  u^4 q^{2 L}+1 -  \left(\gamma  \delta  q^{ K+1}+q^{-K-1}\right) \alpha  \beta  u^2 q^{L}\right).
\end{equation}
Next, let us consider the vectors $\ket{\Bar{u}}$ defined as
\begin{equation}
    \ket{\Bar{u}} = \boldsymbol{B}(\Bar{u},L)\ket{0}, \quad \Bar{u} = \{u_1,u_2, \dots, u_L \},
\end{equation}
where
\begin{equation}
   \boldsymbol{B}(\Bar{u},L) = \mathcal{B}(u_1,L - 1) \mathcal{B}(u_2,L - 2) \dots \mathcal{B}(u_L,0).
\end{equation}
Note that relation \eqref{dor1} implies that $\boldsymbol{B}(\Bar{u},L)$ does not depend on the ordering of the variables $u_i$. Since the vectors $\ket{\bar{u}}$ are obtained by applying $L$ times a tridiagonal matrix on the vector $\ket{0}$, they are contained in the vector space spanned by $\{\ket{0}, \ket{1}, \dots, \ket{L}\}$. As such, they are eigenvectors of $\pi_A$ with eigenvalue $1$, 
\begin{equation}
\pi_A \ket{\bar{u}} = \sum_{i = 0}^L \ket{i}\bra{i}\ket{\bar{u}} = \ket{\bar{u}}.
\end{equation}
The aim is to show that for specific parameters $\Bar{u}$, these vectors are also eigenvectors of $T$. This requires two results. The first is the following relation between the dynamical operator $\mathcal{A}(u,m)$ and product of dynamical operators $\boldsymbol{B}(\Bar{u}, L)$:
\begin{equation}
\begin{split}
    \mathcal{A}(u,m) \boldsymbol{B}(\Bar{u}, L) &= \prod_{i = 1}^L f(u,u_i) \boldsymbol{B}(\Bar{u}, L) \mathcal{A}(u,m-L) \\ 
    & + \sum_{i = 1}^{L} g(u,u_i,m-1)\prod_{\substack{j = 1 \\ i \neq j }}^L f(u_i,u_j) \boldsymbol{B}(\bar u_{\neq i} , u , L )\mathcal{A}(u_i,m-L) \\
    &+ \sum_{i = 1}^{L} w(u,u_i,m-1)\prod_{\substack{j = 1 \\ i \neq j }}^L f(\tau u_i^{-1},u_j)   \boldsymbol{B}(\bar u_{\neq i} , u , L )\mathcal{A}(\tau u_i^{-1},m-L),
\end{split}
\label{be1}
\end{equation}
where
\begin{equation}
    \boldsymbol{B}(\bar u_{\neq i} , u , L ) = \mathcal{B}(u_1,L - 1) \mathcal{B}(u_2,L - 2) \dots   \mathcal{B}(u,m - i) \dots \mathcal{B}(u_L,0).
\end{equation}
This relation is obtained by computing the terms $i =1 $ and by using the symmetry in the indices $u_i$ induced by relation \eqref{dor1}. The second required result is the action of $\mathcal{A}(u,0)$ on the vector $\ket{0}$. From the definition of $\mathcal{A}(u,0)$ in terms of $\boldsymbol{A}$ and $\boldsymbol{A}^*$, and the action \eqref{eq:Hh}-\eqref{ac1} of these operators on the position basis, it follows that
\begin{equation}
        \mathcal{A}(u,0) \ket{0} =  a(u)\ket{0},
        \label{be3}
\end{equation}
where
\begin{equation}
    a(u) = \left( \frac{ q^{-2L}}{(1 - \alpha \beta q) }\left(  \frac{2 \mu_0 \lambda_0 q}{(q+1)}  - \mu_0(\alpha\beta q^2 + 1) +  \frac{q^2 \lambda_0 }{u^2} + \gamma \delta u^2 \lambda_0 \right) + f_1(u,0)\right).
\end{equation}
In particular, we note that this action is diagonal. This feature shows that the \textit{modified} algebraic Bethe ansatz is not necessary and that the model we deal with corresponds to the particular case developed in \cite{cao2003exact}. From this observation and relation \eqref{be1}, it follows that 
\begin{equation}
\begin{split}
T\ket{\bar{u}} &= \Lambda(\bar{u}) \ket{\bar{u}} + \sum_{i = 1}^L E_i(u, \bar{u})  \boldsymbol{B}(\bar u_{\neq i} , u , L ) \ket{0},
\end{split}
\end{equation}
where 
\begin{equation}
\begin{split}
    \Lambda(\Bar{u}) &=  r(u)  a(u)\prod_{i = 1}^L f(u,u_i) +  r(\tau u^{-1}) a(\tau u^{-1}) \prod_{i = 1}^L f(\tau u^{-1},u_i)\\ &
     - \left(r(u) f_1(u,L) + r\left( \tau u^{-1}\right) f_1\left(\tau u^{-1},L\right)\right),
    \end{split}
    \label{eigen1}
\end{equation}
and
\begin{equation}
\begin{split}
   E_i(u,\bar{u}) =  &\Big(r(u) g(u,u_i,L -1)  + r(\tau u^{-1})g(\tau u^{-1},u_i,L -1)\Big) a(u_i)\prod_{\substack{j = 1 \\ i \neq j }}^L f(u_i,u_j) \\
    &  + \Big(r(\tau u^{-1}) w(\tau u^{-1},u_i,L -1) + r(u) w(u,u_i,L -1)\Big)a( \tau u_i^{-1}) \prod_{\substack{j = 1 \\ i \neq j }}^L f(\tau u_i^{-1},u_j).
    \end{split}
    \label{bethe1}
\end{equation}
Thus, for a set of parameters $\Bar{u}$ verifying $E_i(u,\bar{u}) = 0 $, the vector $ \ket{\Bar{u}} = \boldsymbol{B}(\Bar{u},L ) \ket{0}$ is an eigenvector of $T$ with eigenvalues $\Lambda(\bar{u})$, \textit{i.e.}
\begin{equation}
    T \ket{\Bar{u}} = \Lambda(\Bar{u})\ket{\Bar{u}}.
\end{equation}
Since the Heun operator $T$ does not depend on the parameter $u$, the same is true of its eigenvalues $\Lambda(\Bar{u})$. In particular, we find by evaluating \eqref{eigen1} at $u = 0$ that the eigenvalues can be expressed as
\begin{equation}
\begin{split}
    \Lambda(\Bar{u}) &= - (\omega_K + \omega_{K+1})\lambda_L  -\frac{\mu_0 (q-1) }{q }\left(\alpha  \beta q^2+1 + \frac{4 \alpha \beta q^2}{\alpha \beta q - 1}\right) -\frac{(q^2-1)^2 }{q(q+1)} \left( \sum_{i = 1}^L U_i\right),
    \end{split}
    \label{eigen2}
\end{equation} 
where $U_i = \frac{q}{u_i^2} + \alpha \beta \gamma \delta u_i^2 $. Factorizing terms in the variable $u$ in the conditions $ E_i(u,\bar{u}) = 0$, these reduce to the following conditions on $\bar{u}$, referred to as the \textit{Bethe equations},
\begin{equation}
    \begin{split}
       \prod_{\substack{j = 1 \\ i \neq j }}^L \frac{q(u_i^2 - q u_j^2)(\alpha \beta \gamma \delta u_i^2 u_j^2 - 1)}{(u_i^2 - q u_j^2)(\alpha \beta \gamma \delta u_i^2 u_j^2 - q^2)} &=  \frac{  \left(q^2-\alpha  \beta  \gamma  \delta  u_i^4\right) \left(\alpha  \beta  \gamma  \delta  u_i^2 q^{K+L+1}-1\right)}{q^{-2L - 2}\left(\alpha  \beta  \gamma  \delta  u_i^4-1\right) \left(q^{K+L+2}-u_i^2\right)}\times \\ & \quad \frac{(q+1)\left(q^2 u_i^4-1\right) \left(q^{K+1}-\alpha  \beta  u_i^2 q^L\right)  }{ \alpha  \beta    \left(q u_i^4-1\right)  \left(q^L-\gamma  \delta  u_i^2 q^K\right)} \frac{a(u_i)}{a(\tau u_i^{-1})} .
        \label{bethe1}
    \end{split}
\end{equation}
To keep the notation simple, $\bar{u} = \{u_1, u_2, \dots, u_L \}$ will refer from now on to \textit{Bethe roots}, i.e. to solutions of the set of equations \eqref{bethe1}.  
\subsection{Diagonalization of the truncated correlation matrix}

Since the Heun operator $T$ commutes with the truncated correlation matrix and is non-degenerate, its eigenvectors $\ket{\Bar{u}}$ also diagonalize the matrix $C$,
\begin{equation}
    C \ket{\Bar{u}} = c(\Bar{u}) \ket{\Bar{u}}, \quad c(\Bar{u}) \in \mathbb{R}.
\end{equation}
To obtain an explicit expression for the eigenvalues $c(\Bar{u})$ in terms of the parameters $\Bar{u}$, we observe that the action of $\mathcal{B}(u,m)$ in the position basis is tridiagonal and given by
\begin{equation}
\begin{split}
       q^{m+1} \mathcal{B}(u,m)\ket{n} &=  V_{n,m} \ket{n+1} + (X_{n,m} + Y_{n,m} U) \ket{n} + Z_{n,m}\ket{n-1},
\end{split}
\end{equation}
with $U = \frac{q}{u^2} + \alpha \beta \gamma \delta u^2 $. The coefficients $V_{n,m}$, $X_{n,m}$, $Y_{n,m}$ and $Z_{n,m}$ can be computed directly from the definition \eqref{dyop1} and the action of $\boldsymbol{A}$ and $\boldsymbol{A}^*$ in the position basis (see appendix \ref{AppA}). In the case where $\beta = 0$, i.e. the dual $q$-Hahn special case of the $q$-Racah polynomials \cite{koekoek2010hypergeometric}, these coefficients simplify greatly,
\begin{equation} 
    V_{n,m} = \frac{J_n}{q^{n+1}},\quad  X_{n,m} = \alpha  \gamma  q \left(q^{m+1}-q^n\right), \quad     Y_{n,m} =   1- q^{ m + 1-n}, \quad Z_{n,m} =0.
\end{equation}
In particular, $ \mathcal{B}(u,m)$ becomes a raising operator in the sense that $\bra{n-1}\mathcal{B}(u,m) \ket{n} = 0$. This allows to compute the wavefunction $\bra{n}\ket{\Bar{u}}$ of Bethe vectors:
\begin{equation}
    \bra{n}\ket{\Bar{u}} =   {q^{-L(L-1)/2}}\left(\prod_{\ell = 1}^{L-n}1-q^\ell\right) \left(\prod_{i = 0}^{n-1} \frac{J_i}{q^{i+1}}\right) \sum_{r = 0 }^{L-n}\left( \frac{\alpha \gamma q(1-q^{n+1} )}{q - 1}\right)^{L-n-r} S_r(\Bar{U}),
    \label{wf1}
\end{equation}
where $\bar{U} = \{U_1, U_2, \dots, U_L\}$ with $U_i = q u_i^{-2}$. The terms $S_r(\Bar{U})$ are symmetric polynomials of degree $r$ in the variables $U_i$ defined by
\begin{equation}
    S_r(\Bar{U}) = \sum_{i_1 < i_2 < \dots < i_r} U_{i_1}U_{i_2} \dots U_{i_r}, \quad S_0(\Bar{U}) = 1.
\end{equation}
Then, one can use the representation of the truncated correlation matrix in the position basis \eqref{Cposb} to obtain a formula for its eigenvalues in terms of Bethe roots. For any $n \in \{0, 1, \dots L\}$, we find
\begin{equation}
    c(\Bar{u}) = \frac{\bra{n}C\ket{\bar{u}}}{\bra{n}\ket{\bar{u}}} =  \frac{q^{-L(L-1)/2}}{\bra{n}\ket{\Bar{u}}} \sum_{r = 0}^L b_{r,n} S_r(\Bar{U}),
    \label{eigen3}
\end{equation}
where
\begin{equation}
b_{r,n} = \sum_{k = 0}^K \sum_{n' = 0}^{L-r} \phi_n(\omega_k) \phi_{n'}(\omega_k) \left(\prod_{\ell = 1}^{L-n'}1-q^\ell\right) \left(\prod_{i = 0}^{n'-1} \frac{J_i}{q^{i+1}}\right) \left( \frac{\alpha \gamma q(1-q^{n'+1} )}{q - 1}\right)^{L-n'-r}.
\end{equation}
This is valid for parameters $\Bar{u}$ which are solutions of the Bethe equations \eqref{bethe1}. For $\beta =0 $, these equations reduce to
\begin{equation}
     \prod_{\substack{j  = 1 \\ j\neq i}}^L  \frac{\left({u_i^2} -  \frac{u_j^2}{q} \right)}{ \left({u_i^2} - q{u_j^2} \right)} =  \frac{q^{K} \left(q-\alpha  u_i^2\right) \left(q-\gamma  u_i^2\right) \left(q-\gamma  \delta  u_i^2\right)}{ \left(q^{K+L+2}-u_i^2\right) \left(\alpha  \gamma  u_i^2-1\right) \left(\gamma  \delta  u_i^2 q^K-q^L\right)}.
\end{equation}

\section{$TQ$-relations and thermodynamic limit}
\label{s5}
Equations \eqref{eigen2} and \eqref{eigen3} show that the spectra of the Heun operator and of the truncated correlation matrix can be obtained by solving Bethe equations. An alternative approach is given by interpreting the expression \eqref{eigen1} for the eigenvalues of $T$ as a $q$-difference equation. In the case $\beta =0$, \eqref{eigen1} can indeed be rewritten as
\begin{equation}
    \begin{split}
       U^2 Q(U) \Lambda(\Bar{u}) &= \frac{(q+1)(U-\alpha)(U-\gamma)(U-\gamma \delta)}{q^L} Q(qU)- p(U)Q(U)  \\ & + {q^L(q+1)(U - q^{-K-L-1})(U- \alpha \gamma q) (U- \gamma \delta q^{K-L+1})} Q(U/q)
       \label{qdiff}
    \end{split}
\end{equation}
where $p(U)$ is the following polynomial in the variable $U$,
\begin{equation}
\begin{split}
      p(U) &= -\frac{\alpha  \gamma ^2 \delta  (q+1)^2 }{ q^L} + \frac{\gamma  (q+1)U }{q^{K+L}} \left(\alpha  \gamma  \delta  q^{2 K+L+2}+q^K (\alpha  \delta +\alpha +\gamma  \delta +\delta )+\alpha  q^L\right)\\
      & \quad -2 q (\alpha  \gamma +\alpha +\gamma  \delta +\gamma )U^2 +  2 (q+1)U^3,
\end{split}
\end{equation}
and $Q(U)$ is a polynomial of degree $L$, the zeros $U_i$ of which are expressed in terms of entries of a Bethe root $\bar{u} = \{u_1,u_2, \dots u_L\}$:
\begin{equation}
    Q(U) = \prod_{i = 1}^L (U - {U}_i)  = \sum_{i = 1}^L (-1)^{L-i}S_{L-i}(\Bar{U}) U^i.
  \label{Qsym}
\end{equation}
Thus, one can use the zeros of polynomial solutions of equation \eqref{qdiff} to identify Bethe roots. This equation is referred to as the $TQ$-relation in the literature.

Let us now further fix\footnote{The choice of parameters $ \delta = 0$, $ \beta = 0$ corresponds to the affine $q$-Krawtchouk limit of the $q$-Racah polynomials \cite{koekoek2010hypergeometric}. } $\delta = 0$, $\gamma \in [0,1]$ and $q < 1$. Inserting the r.h.s of \eqref{Qsym} in equation \eqref{qdiff} yields a three term recurrence relation for the symmetric polynomials $S_{n}(\bar{U})$:
\begin{equation}
0 = \sigma_{n+1} S_{n+1} + (\rho_{n} + \Lambda(\bar{u})) S_{n}  + \epsilon_{n-1} S_{n-1},
\label{recu1}
\end{equation}
where
\begin{equation}
\sigma_{n} = (q+1) q^{-n}+(q+1) q^{n} -2 (q+1)
\end{equation}
\begin{equation}
\rho_{n} = (q+1) q^{n-L-K} \left(\alpha  \gamma  q^{K+L+1}+\frac{1}{q}\right)+(q+1) (\alpha +\gamma ) q^{-n} -2 q (\alpha  \gamma +\alpha +\gamma )
\end{equation}
\begin{equation}
\epsilon_{n} = -\alpha  \gamma  (q+1) \left(q^{-K}+q^{-L}\right)+ \frac{\alpha  \gamma  (q+1)}{q^{K+L}}  q^{n}+\alpha  \gamma  (q+1) q^{-n}.
\end{equation}
In the thermodynamic limit  $N \rightarrow \infty$, the parameter $\alpha = q^{-N-1}$, with $ 0 < q < 1$, goes to infinity and  \eqref{recu1} becomes effectively a two term recurrence with solution
\begin{equation}
	{S_n}  = {S_L}  \prod_{i =n}^{L-1} \frac{\rho_{i+1} + \Lambda(\bar{u})}{\epsilon_{i}}  + O(\alpha^{-1}).
\end{equation} 
The condition that $S_{-1}(\bar{U}) = 0$ then requires the eigenvalues $\Lambda(\bar{u})$ of $T$ to take certain values

\begin{equation}
 \Lambda(\bar{u}) \in \{-\rho_n + O(\alpha^0) \ |\ n \in \{0,1 \dots, L \} \}.
 \label{approxtherm}
\end{equation}
The spectrum of the Heun operator given by this approximation is compared to spectra found using other methods in Table \ref{t1}. One notes that the values match up to two digits at $N = 49$. This suggests that exact asymptotic results may be obtainable in the thermodynamic limit. 
\begin{table}[h]
\begin{center}
\begin{tabular}{|c | c | c|} 
 \hline
 Solutions of &  $- \rho_n$ for &  Numerical \\
  $S_{-1}(\bar{U}) = 0$ &  $n \in \{0,1,\dots L\}$&  diagonalization of $T$ \\ [0.5ex] 
 \hline\hline
  -778916 & -778741 & -778916 \\ 
 \hline
 -592816 & -592623 & -592816  \\
 \hline
 -444746 & -444544 &  -444746   \\
 \hline
 -327294 &  -327099 &  -327294 \\
 \hline
 -234579 & -234418 &  -234579 \\
 \hline
 -161955 & -161865 &  -161955 \\
 \hline
 -105783 & -105813 &  -105783 \\
 \hline
 -63253.2 & -63460.2 &  -63253.2 \\
 \hline
 -32283.3 & -32687.9 & -32283.6  \\
 \hline
 -11583.9 & -11957.8 & -11583.9 \\
 [1ex] 
 \hline
\end{tabular}
\end{center}
\caption{Eigenvalues of the Heun operator ($N = 49$, $L = 9$, $K = 24$, $q = 0.8$, $\alpha = q^{-N-1}$, $\beta = 0$,  $\gamma = 0.5$, $\delta = 0$) obtained by three methods. The first column are zeros of $S_{-1}(\bar{U})$ seen as a polynomial of degree $L+1$ in $\Lambda(\bar{u})$. The polynomial was obtained by solving the three term recurrence \eqref{recu1}. The second column corresponds to the approximation \eqref{approxtherm} of the spectrum found in the thermodynamic limit. The third is the result of diagonalizing $T$ using \textit{scipy}'s linear algebra package \cite{2020SciPy-NMeth}.}
\label{t1}
\end{table}
\section{Conclusion}

Computing bipartite entanglement for free fermionic chains amounts to determining the spectrum of a truncated correlation matrix. For systems associated to $q$-Racah polynomials, it has been shown how this matrix can be diagonalized via the algebraic Bethe ansatz. In particular, its eigenvalues and eigenvectors have been given in terms of solutions of Bethe equations. The associated Bethe roots were also found to be related to zeros of polynomial solutions of a $q$-difference equation, referred to as the $TQ$-relation. This led to an approximate expression for the eigenvalues of the commuting tridiagonal matrix in the case $\delta = 0$ and $N \rightarrow \infty$. 

While these results do not provide an explicit formula for the bipartite entanglement, it establishes a clear connection between a central problem in quantum many-body physics and a set of tools coming from the study of integrable models. Future research should thus be directed toward investigating, notably in their thermodynamic limit, the solutions of the Bethe equations and $TQ$-relation that were found. Derivation of asymptotic expressions for these would provide the groundwork necessary to analyse the interplay between coupling inhomogeneities in free fermions chains and the presence of entanglement in the ground state.

\section*{Acknowledgements}
We thank Pascal Baseilhac and Rodrigo A. Pimenta for stimulating discussions.
PAB holds an Alexander-Graham-Bell scholarship from the Natural Sciences and
Engineering Research Council of Canada (NSERC). NC is supported by the international research project AAPT of the CNRS and the ANR Project AHA ANR-
18-CE40-0001. The research of LV is founded in part by a Discovery Grant from the Natural Sciences and
Engineering Research Council (NSERC) of Canada.

\begin{appendix}

\section{Appendix}
\label{AppA}

\subsection{$q$-Racah polynomials}

The $q$-Racah polynomials are defined by \cite{koekoek2010hypergeometric}
\begin{equation}
    R_n(\omega_x)=\pFq{4}{3}{q^{-n},\alpha\beta q^{n+1},q^{-x},\gamma\delta q^{x+1}}
    {\alpha q,\beta\delta q,\gamma q }{q;q}
\end{equation}
with 
\begin{equation}
    \omega_x=q^{-x}+\gamma\delta q^{x+1}\,.
\end{equation}
The parameters are restricted by the truncation condition $R_{N+1}(x) = 0$. For instance, one can use $\alpha$ and fix
\begin{equation}
    \alpha = q^{-N-1}.
\end{equation}
These polynomials also satisfy the following recurrence relation
\begin{equation}
    (\omega_x-1-\gamma\delta q) R_n(\omega_x)
    = A_n R_{n+1}(\omega_x)-(A_n+C_n)R_n(\omega_x)+C_n R_{n-1}(\omega_x)
    \label{recurel}
\end{equation}
where
\begin{eqnarray}\label{eq:A}
    A_n&=& \frac{\left(\alpha  q^{n+1}-1\right)\left(\gamma  q^{n+1}-1\right) \left(\alpha  \beta  q^{n+1}-1\right) \left(\beta  \delta q^{n+1}-1\right)}
   {(1-\alpha  \beta  q^{2 n+1})(1-\alpha  \beta  q^{2 n+2})}\\
   \label{eq:C}
    C_n&=& \frac{ \left(\beta 
   q^n-1\right) \left(\alpha  q^n-\delta \right)
   \left(\alpha  \beta  q^n-\gamma \right)(q^{
   n+1}-q)}
   {(1-\alpha  \beta  q^{2 n})(1-\alpha 
   \beta  q^{2 n+1})}
\end{eqnarray}
The normalisation weight is
\begin{equation}
    W_{k} = \frac{(\beta^{-1}\gamma q, \delta q; q)_N(\gamma\delta q, \alpha q, \beta \delta q, \gamma q ; q)_k (1 - \gamma \delta q^{2k+1})}{(\gamma \delta q^2, \beta^{-1}; q)_N(q,\alpha^{-1}\gamma \delta q, \beta^{-1}\gamma q, \delta q ; q)_k (\alpha \beta q)^k (1- \gamma \delta q)}.
\end{equation}
These polynomials also have a difference equation of the form \eqref{tri2}, with coefficients given by
\begin{equation}
\begin{split}
    \Bar{J}_k &= \sqrt{\frac{(1-\alpha q^{k+1})(1-\beta \delta q^{k+1})(1-\gamma q^{k+1})(1-\gamma \delta q^{k+1}) }{(1-\gamma \delta q^{2k+1})(1-\gamma \delta q^{2k+2})}}\times \\
    & \sqrt{\frac{(1- q^{k+1})(1-\alpha^{-1}\gamma \delta q^{k+1})(1-\beta^{-1}\gamma q^{k+1})(1-\delta q^{k+1})(\alpha \beta q)}{(1-\gamma \delta q^{2k+2})(1-\gamma \delta q^{2k+3})}}
\end{split} 
 \end{equation}
 \begin{equation}
 \begin{split}
     \Bar{\mu}_k &= \frac{(1-\alpha q^{k+1})(1-\beta \delta q^{k+1})(1-\gamma q^{k+1})(1-\gamma \delta q^{k+1})}{(1-\gamma \delta q^{2k+1})(1-\gamma \delta q^{2k+2})} \\ &+ \frac{q (1- q^{k})(1- \delta q^{k})(\beta-\gamma q^{k})(\alpha-\gamma \delta q^{k})}{(1-\gamma \delta q^{2k})(1-\gamma \delta q^{2k+1})} -1 - \alpha \beta q
     \end{split}
 \end{equation}
 
\subsection{Functions in the algebraic Bethe ansatz}

The functions in the definition of the dynamical operators are:
 \begin{equation}
 \begin{split}
     f_1(u,m) &= \frac{2 q^{m+1} \left(q+\alpha  \beta  \gamma  \delta  u^4\right)}{u^2 \left(\alpha  \beta  q^{2 L+2 m+1}-q^{2 L}\right)} -\frac{  u^2 \eta (q+ 1) q^{-2L+2}}{(q^2-1)^2 \left(q^2-\alpha  \beta  \gamma  \delta  u^4\right)} \\
     & +  \frac{(q+1)}{(q^2 - 1)^2\left(q^{2 L}-\alpha  \beta  q^{2 L+2 m+1}\right)} \left(\frac{\eta^*(\gamma  \delta   u^4 q^{-2L+1}- q^{2 m+4})}{ \left(q^2-\alpha  \beta  \gamma  \delta  u^4\right)} - {2 \xi  q^{m+2}} \right),
      \end{split}
 \end{equation}
 and
 \begin{equation}
 \begin{split}
      f_2(u,m) &= \frac{  \left(\alpha  \beta  q^{2 L+2 m+3}+q^{2 L}\right) \left(q+\alpha  \beta  \gamma  \delta  u^4\right)}{u^2q^{m+2 L+1}} + \frac{ \left(\eta^* q^{2L+m+1}+\alpha  \beta  \xi  q^{2 L+2 m+3}+\xi  q^{2 L}\right)}{  q^{m+2 L}(q-1)^2 (q+1)}.
 \end{split}
 \end{equation}
 The functions in the relation between the dynamical operators are:
 
 \begin{equation}
    f(u,v) = \frac{\left(u^2-q v^2\right) \left(\alpha  \beta  \gamma  \delta  u^2 v^2-q^2\right)}{q \left(u^2-v^2\right) \left(\alpha  \beta  \gamma  \delta  u^2 v^2-q\right)},
\end{equation}
\begin{equation}
    g(u,v,m) = \frac{(q-1) \left(q^2-\alpha  \beta  \gamma  \delta  v^4\right) \left(\alpha  \beta  v^2 q^{2 L+2 m+3}-u^2 q^{2 L}\right)}{q \left(u^2-v^2\right) \left(\alpha  \beta  q^{2 L+2 m+3}-q^{2 L}\right) \left(q-\alpha  \beta  \gamma  \delta  v^4\right)},
\end{equation}
and
\begin{equation}
    w(u,v,m) =  \frac{\alpha  \beta  (q-1) \left(\alpha  \beta  \gamma  \delta  v^4-1\right) \left(\gamma  \delta  u^2 v^2 q^{2 L}-q^{2 (L+m+2)}\right)}{\left(q^{2 L}-\alpha  \beta  q^{2 L+2 m+3}\right) \left(q-\alpha  \beta  \gamma  \delta  v^4\right) \left(q-\alpha  \beta  \gamma  \delta  u^2 v^2\right)}.
\end{equation}
The coefficients giving the action of $\mathcal{B}(u,m)$ on vectors in the position basis are:
\begin{equation} 
    V_{n,m} = J_n\left( q^{-n-1} - \alpha \beta q^{  m + 2}  - \alpha \beta q^{  m + 1} +  \alpha^2 \beta^2 q^{ 2m +n + 4}\right),
\end{equation}
\begin{equation}
    X_{n,m} = -\frac{ \mu_n \lambda_n (\alpha \beta q^{2m + 3} + 1)}{(q+1)} - \frac{\alpha \beta (q+1)}{q^{ - m - 1}}  \mu_n + \frac{(\eta^* q^{2L + m  + 1} + \alpha \beta \xi q^{2L + 2m + 3} +  \xi q^{2L})}{(q-1)^2 (q+1) q^{ 2L - 1}},
\end{equation}
\begin{equation}
    Y_{n,m} = - q^{  m + 1} \lambda_n + \frac{(\alpha \beta q^{2L + 2m + 3} + q^{2L})}{q^{ 2L}},
\end{equation}
and
\begin{equation}
        Z_{n,m} = J_{n-1} \alpha \beta \left( q^{ 2m+ 3 - n} + q^n - q^{m + 2} - q^{ m + 1}\right).
\end{equation}

\end{appendix}

% TODO:
% Provide your bibliography here. You have two options:

% FIRST OPTION - write your entries here directly, following the example below, including Author(s), Title, Journal Ref. with year in parentheses at the end, followed by the DOI number.
%\begin{thebibliography}{99}
%\bibitem{1931_Bethe_ZP_71} H. A. Bethe, {\it Zur Theorie der Metalle. i. Eigenwerte und Eigenfunktionen der linearen Atomkette}, Zeit. f{\"u}r Phys. {\bf 71}, 205 (1931), \doi{10.1007\%2FBF01341708}.
%\bibitem{arXiv:1108.2700} P. Ginsparg, {\it It was twenty years ago today... }, \url{http://arxiv.org/abs/1108.2700}.
%\end{thebibliography}

% SECOND OPTION:
% Use your bibtex library
\nolinenumbers

\end{document}